\documentclass{aa}  
\usepackage{graphicx}
\usepackage{txfonts}
\begin{document}
   \title{Predictions of polarized dust emission from interstellar clouds: spatial variations in the efficiency of radiative torque alignment}
   \titlerunning{Predictions of polarized dust emission from interstellar clouds}

   \author{V.-M. Pelkonen
          \inst{1}
          \and
          M. Juvela\inst{1}
	  \and
	  P. Padoan \inst{2} 
          }


   \institute{Observatory, University of Helsinki,
              T\"ahtitorninm\"aki, P.O. Box 14,
	      FI-00014 University of Helsinki, Finland \\
	      \email{veli-matti.pelkonen@.helsinki.fi, mika.juvela@helsinki.fi}
	     \and
	     Department of Physics, University of California, San Diego,
	     CASS/UCSD 0424, 9500 Gilman Drive, La Jolla, CA 92093-0424 \\
	     \email{ppadoan@ucsd.edu}}


   \date{}

 
  \abstract
   {Polarization carries information about the magnetic fields in
   interstellar clouds. The observations of polarized dust emission are used to study the role of magnetic
   fields in the evolution of molecular clouds and the initial phases of
   star formation. Therefore, it is important to understand how different cloud
   regions contribute to the observed polarized signal.}
   {We study the grain alignment with realistic simulations, assuming the 
   radiative torques to be the main mechanism. The
   aim is to study the efficiency of the grain alignment as a function of cloud
   position and to study the observable consequences of these spatial variations.}
   {Our results are based on the analysis of model clouds derived from MHD
   simulations of super-Alfv\'enic magnetized turbulent flows. The continuum
   radiative transfer problem is solved with Monte Carlo methods to estimate
   the three-dimensional distribution of dust emission and the radiation
   field strength. The anisotropy of the
   radiation field is taken into account explicitly. We also examine the effect of grain growth in cores both to the observed polarization and to the inferred magnetic field.}
   {Using the assumptions of Cho \& Lazarian, our findings are generally consistent with their results. However, the anisotropy factor is lower than their assumption of $\gamma = 0.7$, and thus radiative torques are less efficient. Compared with our previous paper, $P/I$ relations are steeper. Without grain growth, the magnetic field of the cores is poorly recovered above a few $A_{\rm V}$. If grain size is doubled, the polarized dust emission can trace the magnetic field lines possibly up to $A_{\rm V} \sim 10$ magnitudes. However, many of the prestellar cores may be too young for grain coagulation to play a major role. The inclusion of direction-dependent radiative torque efficiency weakens the alignment. Even with doubled grain size, we would not expect to probe the magnetic field beyond a few magnitudes in $A_{\rm V}$.}
   {}


   \maketitle
%

\section{Introduction}

Magnetic fields are important in many astrophysical processes. In star
formation, the magnetic pressure may support an otherwise gravitationally unstable cloud core
against collapse. Magnetohydrodynamical models allow predictions of what
magnetic fields might look like in interstellar clouds and protostellar cores,
but the models need to be tested against observations. Information about
interstellar magnetic fields can be derived with different techniques.
Amongst the most useful are the polarization of starlight from background
stars and the polarized thermal dust emission at longer wavelengths, which
arise from the alignment of the spin axis of the dust grains along the
magnetic field.

The main question with dust polarization is the alignment mechanism, or
rather, what makes the dust grains spin up in the first place? One of the
earliest mechanisms proposed was the paramagnetic mechanism (Davis \&
Greenstein \cite{davisgreen}), which is based on the direct interaction of
rotating grains with the interstellar magnetic field. However, to explain the
grain alignment needed for the polarization, this mechanism
requires much stronger magnetic fields than have been observed. Purcell (\cite{purcell})
introduced several processes that could cause the grains to become very fast rotators,
and suggested that ${\rm H_2}$-ejection might be a major cause of fast grain
rotation. However, further investigations of ${\rm H_2}$-ejection identified
several related processes that make it inefficient in aligning dust
grains, such as grain wobbling (e.g., Jones \& Spitzer \cite{josp67}; Lazarian
\cite{la94}; Lazarian \& Roberge \cite{laro97}), grain flipping (Lazarian \&
Draine \cite{ladr99a}), and ``nuclear relaxation'' (Lazarian \& Draine
\cite{ladr99b}). These mechanisms and others were discussed in a review
paper by Lazarian (\cite{lazarian}).

The radiative torque mechanism has become a strong
candidate for the primary mechanism of grain alignment inside molecular clouds.
The radiative torque mechanism involves the transfer of momentum by collisions of
photons onto the grain, causing a torque that rotates the grain around
its axis. It was first introduced by Dolginov (\cite{do72}) and Dolginov \&
Mytrophanov (\cite{domy76}). The efficiency of the radiative torques has been
demonstrated using numerical simulations (Draine \& Weingartner \cite{drwe96},
 \cite{drwe97}) and in a laboratory setup (Abbas et al. \cite{abal04}). The 
predictions of the radiative torque mechanisms are roughly consistent with 
observations (e.g., Lazarian et al. \cite{laal97}; Hildebrand et al. 
\cite{hial00}).

However, Draine \& Weingartner (\cite{drwe96}) were incorrect to assume 
that the grains would just be spun up by radiative torques, 
and aligned by paramagnetic relaxation. In their subsequent paper, Draine \& Weingartner (\cite{drwe97}) confirmed that radiative torques would be able to align grains on their own, without relying on paramagnetic relaxation. Lazarian \& Draine (\cite{ladr99a}) showed that thermal flipping would prevent the paramagnetic alignment mechanism being effective in small grains.

Cho \& Lazarian (\cite{cholazarian}) used a spherically symmetric model cloud
to calculate the size of grains aligned by radiative torques, assuming a
constant anisotropy factor, $\gamma = 0.7$, and neglecting the isotropic
component of the radiation. Their calculations showed that even deep inside
GMCs ($A_{\rm V} \leq 10$), large grains can still be aligned by radiative torques.
Cho \& Lazarian (\cite{cholazarian}) presented an empirical formula for the
minimum size of the aligned grain $a_{\rm alg}$ as a function of the density $n_H$
and extinction $A_{\rm V}$, which is
\begin{eqnarray}
a_{{\rm alg}} = (log \;n_{\rm H})^3(A_{\rm V, 1D} + 5) / 2800 \; \mu{\rm m}.
\label{eqaalg}
\end{eqnarray}
The formula is, strictly speaking, only valid for spherically symmetric clouds, and
carries the additional assumption of a constant anisotropy factor, $\gamma =
0.7$. Pelkonen et al. (\cite{Pelkonen2007}) used this formula to model
polarization in a clumpy cloud.

The obvious next step is to complete radiative transfer calculations in more detail, derive the actual anisotropy within an inhomogeneous cloud, and use it to
calculate the efficiency of radiative torques. Similar work was carried out by Bethell et al. (\cite{bethell}). These authors were able to derive $P/I$ relations reminiscent of the observed ones, and their results agreed with the qualitative results of Cho \& Lazarian (\cite{cholazarian}). They found that simplifying assumptions about the radiative anisotropy (e.g., a constant value of $\gamma$) have only a moderate effect on the emergent degree of polarization.

In the present paper, we study the anisotropy factor $\gamma$ and calculate the minimum sizes of the
aligned grains within our model clouds. In addition to calculating 1D models
to compare with Eq. 1, we also compare our results with those of our previous 3D
models (Pelkonen et al. \cite{Pelkonen2007}). We investigate quantitatively
the effects that grain growth has on the polarized signal expected from dense
cloud cores. In particular, we examine a model where the grain growth is restricted to the densest regions. Furthermore, we examine where the observed polarization actually originates along the line of sight.

In the past few years, our understanding of the radiative torques has been improving rapidly. Lazarian \& Hoang (\cite{Lazarian2007}) presented the first analytic model of radiative torques, allowing for quantitative treatment of radiative torque alignment. At the time of submission of this paper, Hoang \& Lazarian (\cite{hola08}) expanded their treatment of radiative torques from beam radiation fields to dipole and quadrupole fields, and illustrated the dependence of radiative torque efficiency on the angle between the magnetic field and the radiation. These important steps should be included in future studies of polarization within clumpy clouds. In this paper, we present an example where we take into account the angle between the local magnetic field and the direction of the anisotropic beam radiation to quantify the effect that it will have on emergent polarization.

\begin{figure}
\centering 
\includegraphics[width=8cm]{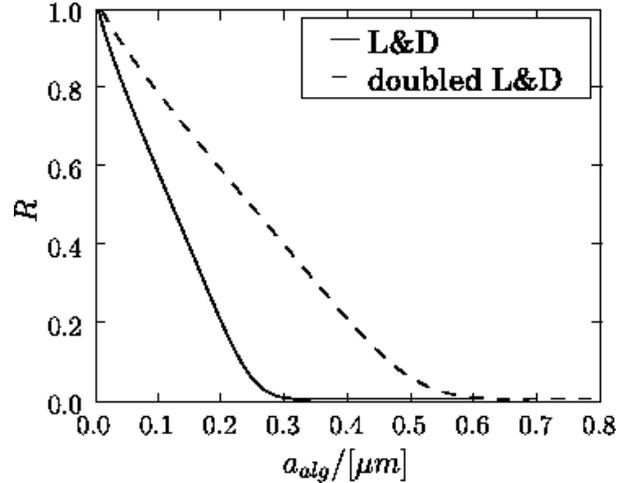} 
\caption{
Polarization reduction factor $R$ as a function of $a_{\rm alg}$ using the
 normal ({\it solid line}) and doubled ({\it dashed line}) Li \& Draine
(\cite{Li2001}) grain sizes.
} 
\label{Raalg}

\end{figure}
\begin{figure}
\centering 
\includegraphics[width=8cm]{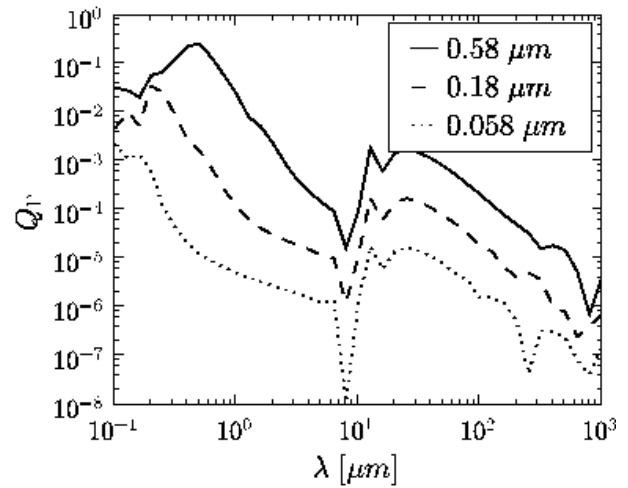} 
\caption{
Radiative torque efficiency $Q_{\Gamma}$ for different grain sizes, $a$, as a function of $\lambda$ (J. Cho, private communication).
} 
\label{fig2}
\end{figure}

\begin{figure*}
\centering 
\includegraphics[width=16cm]{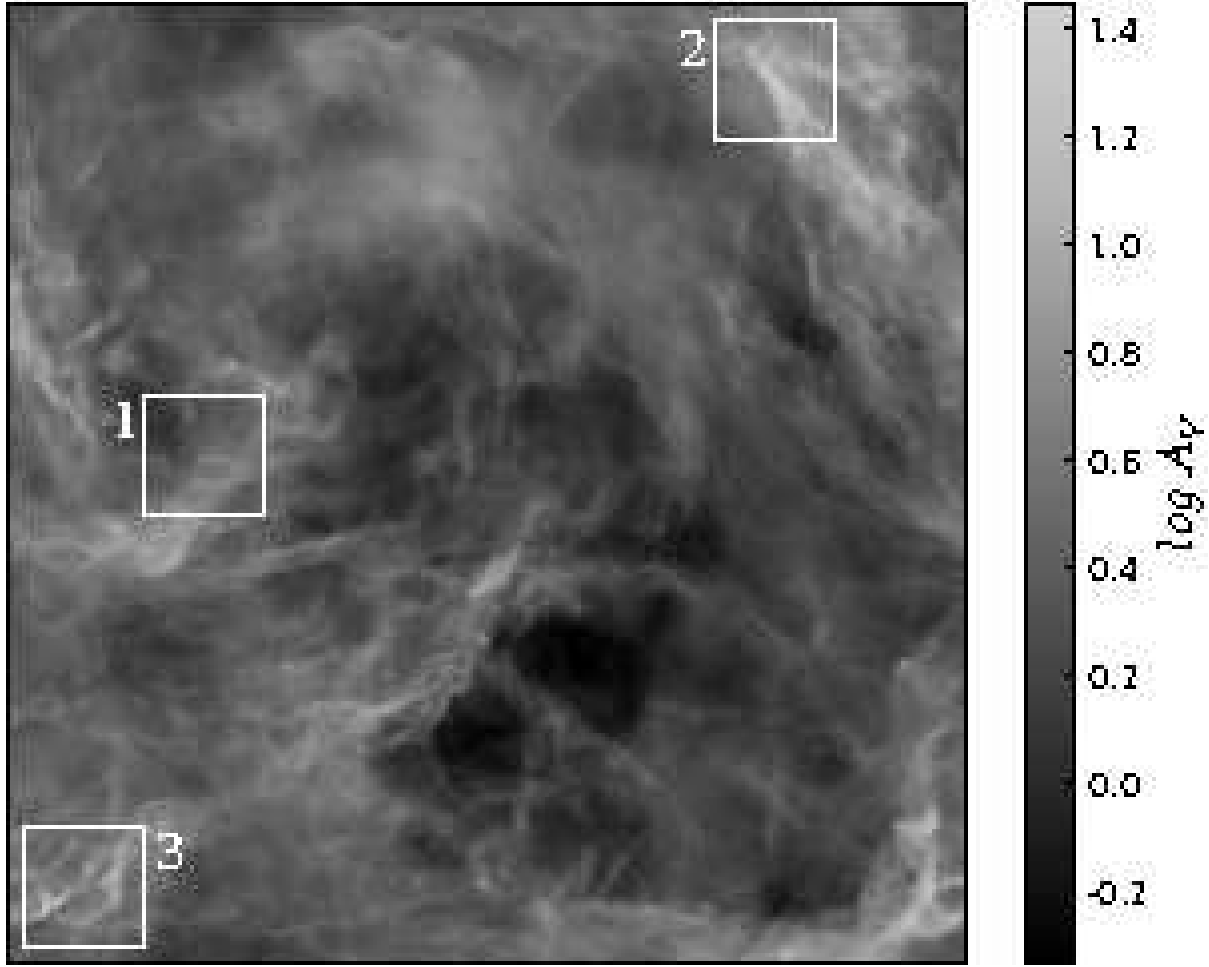} 
\caption{
$A_{\rm V}$ map of the high resolution 3D model as seen from the $z$-direction. The selected core regions are marked with white boxes. 
} 
\label{fig:1024_cloud_projection}
\end{figure*}

\section{The basic equations}

To calculate the alignment efficiency of radiative torques, we follow the formalism presented in Draine \& Weingartner (\cite{drwe96}). The rotational
damping time due to gas drag is
\begin{eqnarray}
\tau_{\rm drag, gas} & = & \frac {\pi \alpha_1 \rho a_{\rm eff}}{3 \delta n_{\rm H} ( 2 \pi m_{\rm H} k T)^{1/2}} = (8.74 \times 10^4 \; {\rm yr}) \nonumber \\
 & & \times \frac{\alpha_1}{\delta} \rho_3 a_{-5} T_{2}^{1/2} ( \frac{3000 \; {\rm cm^{-3} \; K}}{n_{\rm H} T}),
\end{eqnarray}
where $\alpha_1$ is a geometric factor related to the moment of inertia (see
Eq. 38 and Table 2 in Draine \& Weingartner \cite{drwe96}), $\delta \approx
\alpha_1$, $\rho$ is the solid density, $\rho_3 = \rho/3 \; {\rm g \;
cm^{-3}}$, $a_{\rm eff}$ is the radius of a sphere of an equal volume, $a_{-5} =
a_{\rm eff}/10^{-5} \; {\rm cm}$, $n_{\rm H}$ is the number density of hydrogen
atoms, $T$ is the gas temperature, and $T_{2} = T/10^2 {\rm \; K}$. The
rotational dampening time due to thermal emission of photons by a grain heated
by starlight to a temperature $T_{\rm d}$ is
\begin{eqnarray}
\tau_{\rm drag, em} = (1.60 \times 10^5 \; {\rm yr}) \frac{\alpha_1 \rho_3 a_{-5}^3}{\langle Q_{\rm abs} \rangle} (\frac{T_{\rm d}}{18 \; {\rm K}})^2 (\frac{u_{\rm ISRF}}{u_{\rm rad}}),
\label{eq1}
\end{eqnarray}
where $u_{\rm ISRF}$ and $u_{\rm rad}$ are the energy density of the interstellar
radiation field outside the cloud (Mathis et al. \cite{maal83}) and the radiation field
illuminating the grain, respectively, and
\begin{eqnarray}
\langle Q_{\rm abs} \rangle = \frac{1}{u_{\rm rad}} \int u_{\lambda} Q_{\rm abs}(\lambda) \; d\lambda,
\label{eq2}
\end{eqnarray}
and $Q_{\rm abs} \propto \lambda^{-2}$.

If we assume that the grain is illuminated by an isotropic radiation component
$u_{\rm rad}^{\rm iso}$ and an anisotropic radiation component $u_{\rm rad}^{\rm ani}$, a steady
radiative torque gives the grain an angular velocity (Eq.65 in Draine \& Weingartner \cite{drwe96})
\begin{eqnarray}
\omega_{\rm rad} & = & \frac{5 \bar{\lambda}}{8 \delta a_{\rm eff}^2} (\frac{kT}{8\pi m_{\rm H}}){1/2}(\frac{1}{n_{\rm H} kT}) \nonumber\\
 & & \times (u_{\rm rad}^{\rm iso} \langle Q_{\Gamma}^{\rm iso} \rangle + u_{\rm rad}^{\rm ani} \langle {\bf Q_{\Gamma}} \rangle \cdot {\bf \hat{a}_1}) (\frac {1}{1 + \frac{\tau_{\rm drag,gas}}{\tau_{\rm drag,em}}}),
\label{eq3}
\end{eqnarray}
where $\bar{\lambda}$ is the average wavelength, $m_{\rm H}$ the mass of a
hydrogen atom, $\langle Q_{\Gamma}^{\rm iso} \rangle$ and $\langle {\bf Q_{\Gamma}} \rangle$ are the isotropic and anisotropic radiative torque
efficiency, respectively, and ${\bf \hat{a}_1}$ is the unit vector of the
rotational axis. However, this is an approximate treatment, since we ignore the dependence of radiative torque efficiency on the angle between the magnetic field and the radiation. Radiative torque efficiency is lower by more than a factor of ten for angles close to being perpendicular.

The randomization of the rotation of the grain is caused by
collisions with gas molecules. When the grain rotates much more rapidly than the thermal rotation rate,
\begin{eqnarray}
\omega_{T}^2 = \frac{15kT}{8\pi \alpha_1 \rho a_{eff}^5},
\label{eq4}
\end{eqnarray}
this randomization is greatly reduced. Thus, a suprathermally rotating grain
is expected to be aligned with the magnetic field. The factor between the
rotational rate due to the radiative torques, $\omega_{\rm rad}$, and the thermal
rotational rate, $\omega_T$, is an indicator of the efficiency of the grain
alignment, given by
\begin{eqnarray}
(\frac{\omega_{\rm rad}}{\omega_{T}})^2 & = & \frac{5 \alpha_1}{192 \delta^2} \frac{\rho a_{\rm eff}}{m_{\rm H}} (\frac{1}{n_H k T})^2  \nonumber \\
 & & \times \lbrack \int{d\lambda ({\bf Q_{\Gamma}}  \cdot {\bf \hat{a}_1}) \lambda (4 \pi J_{\lambda}/c)} \rbrack^2 (\frac {1}{1 + \frac{\tau_{\rm drag,gas}}{\tau_{\rm drag,em}}})^2.
\label{eq5}
\end{eqnarray}

\begin{figure}
\centering 
\includegraphics[width=8cm]{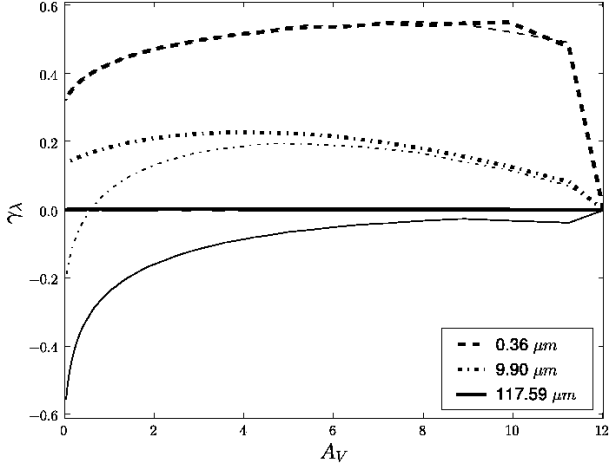} 
\caption{
The anisotropy factors at different wavelengths inside a 1D model cloud. The thick lines are without and the thin lines with dust emission.
} 
\label{1d_aniso_Av}
\end{figure}

\begin{figure}
\centering 
\includegraphics[width=8cm]{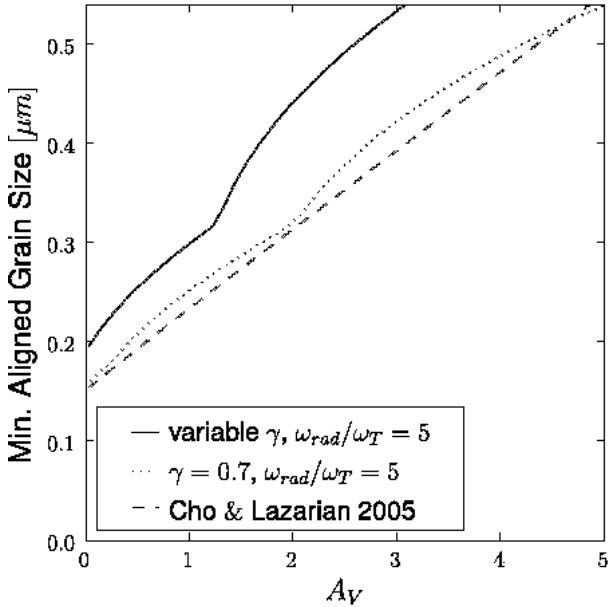} 
\caption{
Comparison of Eq.~\ref{eqaalg} ({\it dashed}) and our results ({\it solid, dotted}) for the Bonnor-Ebert sphere model. 
} 
\label{1d_min_align}
\end{figure}

\section{Polarization}
\subsection{Rayleigh polarization reduction factor}

The Rayleigh polarization reduction factor $R$ is a measure of the imperfect
alignment of dust grains with respect to the magnetic field (Greenberg
\cite{greenberg}; see also Lee \& Draine \cite{leedraine}). The degree of
polarization is reduced when grains are not aligned with the magnetic
field. In the case of radiative torques, smaller grains are not aligned but
larger grains are. The minimum aligned grain size, $a_{\rm alg}$, is
derived from Eq.~\ref{eq5}, such that $a_{\rm eff} = a_{\rm alg}$ when
$\omega_{\rm rad}/\omega_{T} > 5$.

The polarization reduction factor is
\begin{eqnarray}
R = \frac{\int_{a_{{\rm alg}}}^{a_{{\rm max}}} \; C_{{\rm ran}}n(a) \; da}{\int_{a_{{\rm min}}}^{a_{{\rm max}}} \; C_{{\rm ran}}n(a) \; da},
\label{eqR}
\end{eqnarray}
where $C_{{\rm ran}}$ is the polarization cross section of a randomly aligned
grain, $n(a)$ the grain number density, $a$ the grain size, $a_{{\rm min}}$
the minimum size of the grains, and $a_{{\rm max}}$ the maximum size. This
integral is, of course, sensitive to the limits and the form of the grain size distribution (see Fig.~\ref{Raalg}). In the
following, we use $a_{{\rm min}} = 0.005 \; {\rm \mu m}$ and the  Li \& Draine
(\cite{Li2001}) grain size distribution.

\subsection{Polarized thermal dust emission}

The polarized thermal dust emission is calculated following the formalism in
Fiege \& Pudritz (\cite{fipu00}). Self-absorption and scattering can be
neglected at submillimeter wavelengths. The Stokes Q and U components are
equal to the integrals

\begin{eqnarray}
q & = & \int \alpha I_{\lambda} \cos 2\psi \cos^2 \gamma \; ds \;,\label{eqq}\\
u & = & \int \alpha I_{\lambda} \sin 2\psi \cos^2 \gamma \; ds \;,
\label{equ}
\end{eqnarray}
where $\alpha$ is a coefficient of the particle properties to be defined
later, $\psi$ the angle between the projection of $\bf{B}$ on the plane of the
sky and the north, and $\gamma$ the angle between the local $\bf{B}$ vector
and the plane of the sky. The dust emission intensity at the given wavelength,
$I_{\lambda}$, is obtained from radiative transfer calculations.

The polarization angle $\chi$ is given by
\begin{eqnarray}
\tan 2\chi= \frac{u}{q},
\end{eqnarray}
and the degree of polarization $P$ is 
\begin{eqnarray}
P = \frac {\sqrt{q^2 + u^2}}{\Sigma - \Sigma_2} \;,
\label{eqP}
\end{eqnarray}
with
\begin{eqnarray}
\Sigma = \int I_{\lambda} \; ds \;,
\end{eqnarray}
and 
\begin{eqnarray}
\Sigma_2 = \frac{1}{2} \int \alpha I_{\lambda} (\cos^2 \gamma - \frac{2}{3}) \; ds \;,
\end{eqnarray}
where $\Sigma$ is the intensity of the dust emission along the line of sight, and $\Sigma_2$ is a quantity related to that, which is determined by the extinction cross section (see Eq.10 in Fiege \& Pudritz \cite{fipu00}).

\begin{figure*}
\centering 
\includegraphics[width=16cm]{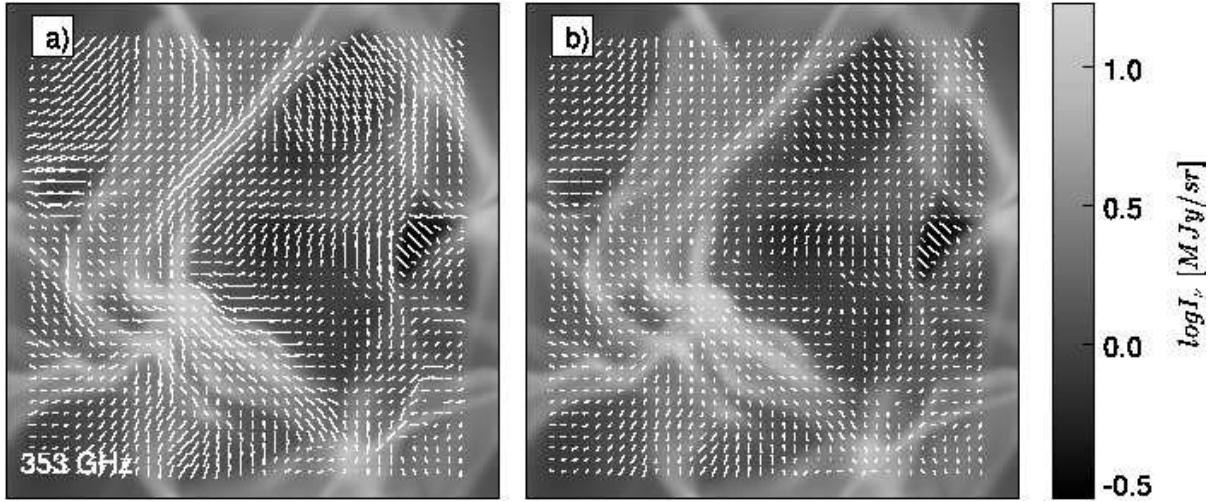} 
\caption{
Simulated polarization maps at 353 GHz with $R$ calculated using Eq.~\ref{eqaalg} (left frame) and Eq.~\ref{eq5}. Polarization vectors are drawn for every third pixel and the scaling of the vector lengths is the same in both frames, representing the polarization degree. In the left frame the maximum polarization degree is 8.5\%, while in the right frame it is 7.3\%. The background image shows the logarithm of the total intensity at this frequency.
} 
\label{old_new_emitted_353}
\end{figure*}

The coefficient $\alpha$ is defined as
\begin{eqnarray}
\alpha = R F \frac{C_{\rm pol}}{C_{\rm ran}},
\label{eqalpha}
\end{eqnarray}
where $R$ is the Rayleigh polarization reduction factor due to imperfect grain
alignment, $F$ is the polarization factor due to the turbulent component of
the magnetic field, $C_{\rm pol}$ is the grain polarization cross section, and
$C_{\rm ran}$ is the average cross section of a randomly aligned grain. The polarization cross section of an oblate grain is the difference between the main axis cross section and the minor axis cross sections of the grain (for more details, see Lee \& Draine \cite{leedraine}). 

In our study, $F = 1$ because the three-dimensional magnetic field is given by
the MHD models, and we assume that the small-scale structure is resolved in
the numerical solution. The ratio $C_{\rm pol}/C_{\rm ran}$ needs to be fixed
because we want to study the effects caused by variations in $R$.  To avoid
an unreasonably high polarization degree, we choose $C_{\rm pol}/C_{\rm ran} =
0.15$. This corresponds to the axial ratio of roughly 1.1 (see Fig. 1 in
Padoan et al. \cite{paal01}).

\section{Radiative transfer calculations}

To calculate the amount of polarized dust emission, we needed to carry
out radiative transfer modeling, to (1) calculate the intensity and angular distribution of the radiation field, (2) determine the resulting total intensity of dust emission, and (3) determine the dust temperature. Apart from determining the total emission, dust temperatures affect thermal rotation rates and, thereby, the minimum size of aligned grains. The intensity of the incoming radiation and its anisotropy affect the magnitude of the rotational torques, as shown in Eq.~\ref{eq5}.

The radiative transfer calculations are made with a Monte Carlo program
(Juvela \& Padoan \cite{Juvela2003}), where the contribution of transiently heated,
small grains is also solved. To determine the anisotropy of the
radiation field within the model clouds, the photons entering each
computational cell were registered. The angular distribution was tracked by
binning the photons into 12 bins that corresponded to angular discretization
according to the Healpix scheme (Gorski et al. \cite{Gorski}). The anisotropy of the radiation field
was estimated by calculating the vector sum of the intensity over different
directions. We made the simplifying assumption that the radiative torque
efficiency is direction invariant and that opposing directions cancel out. The
remaining part was our anisotropic radiation component. When the anisotropic
component was subtracted from the total radiation, we were left with the
isotropic component. Based on Table 4 in Draine \& Weingartner (\cite{drwe96}), we assumed that $Q_{\Gamma}^{\rm iso} = 0.1 \times Q_{\Gamma}$, where $Q_{\Gamma}$ is the radiative torque efficiency for anisotropic radiation, shown in Fig.~\ref{fig2}. We used the dust model discussed in Li \& Draine (\cite{Li2001}).

Finally, we calculated an example where we took the direction dependence of the radiative torque efficiency from Fig. 17 in Hoang \& Lazarian (\cite{hola08}). We normalized the unidirectional curve to 1 when the radiation vector and the magnetic field vector are parallel, and used the resulting curve as our direction-dependent efficiency factor. The efficiency factor drops by more than a factor of ten when moving towards a perpendicular orientation of the vectors. We considered neither multipoles nor isotropy in this simplified example, but multiplied each beam direction by the efficiency factor, and calculated the vector sum of our anisotropy. This anisotropic radiation intensity was then used to calculate the aligned grain size.

\begin{figure*}
\centering 
\includegraphics[width=16cm]{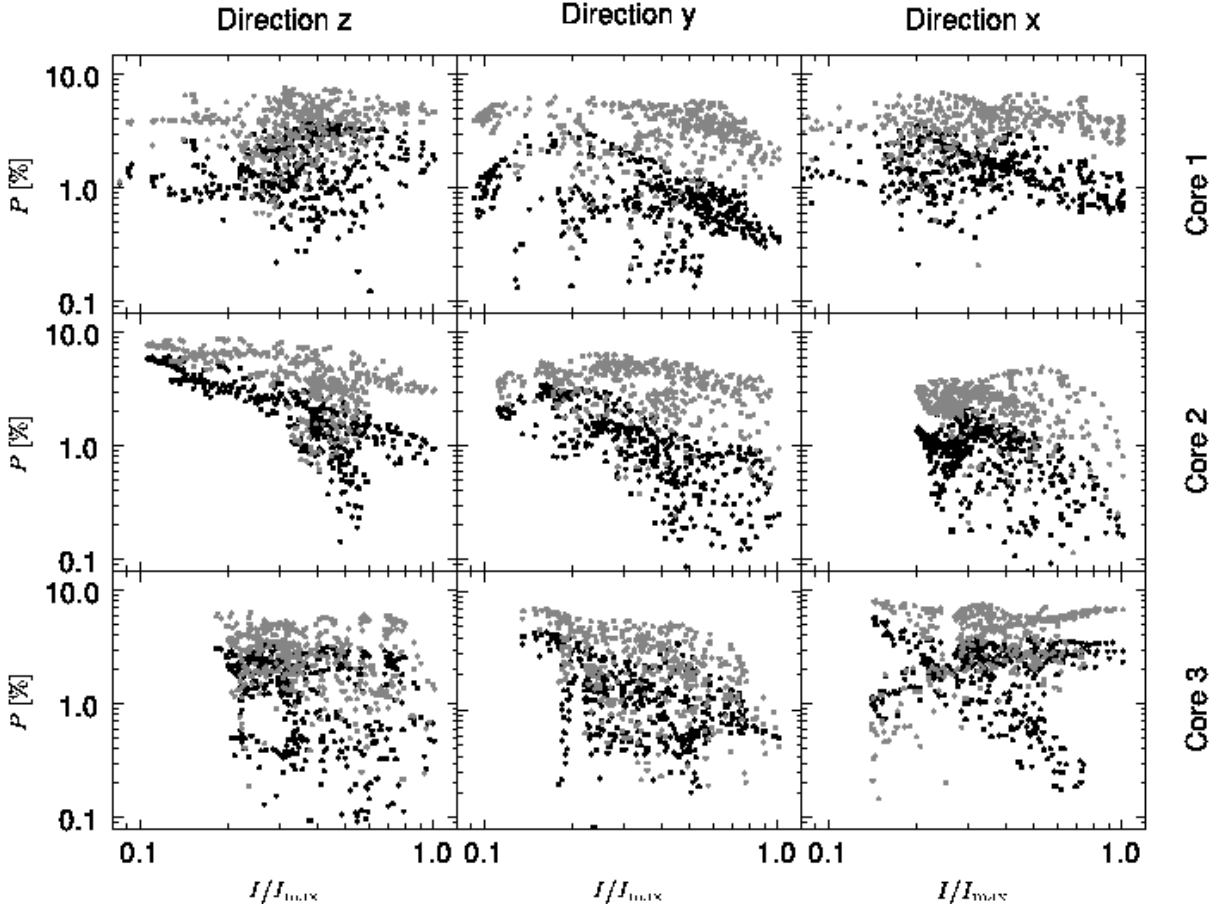} 
\caption{
The relation between polarization degree and total intensity for selected cores of the polarization maps in Fig.~\ref{old_new_emitted_353}. Cyan dots show the old results (Eq.~\ref{eqaalg}), assuming constant $\gamma$, while black dots show the new ones (Eq.~\ref{eq5}) with spatially varying $\gamma$. 
} 
\label{old_new_corepoldeg}
\end{figure*}

\begin{figure*}
\centering 
\includegraphics[width=16cm]{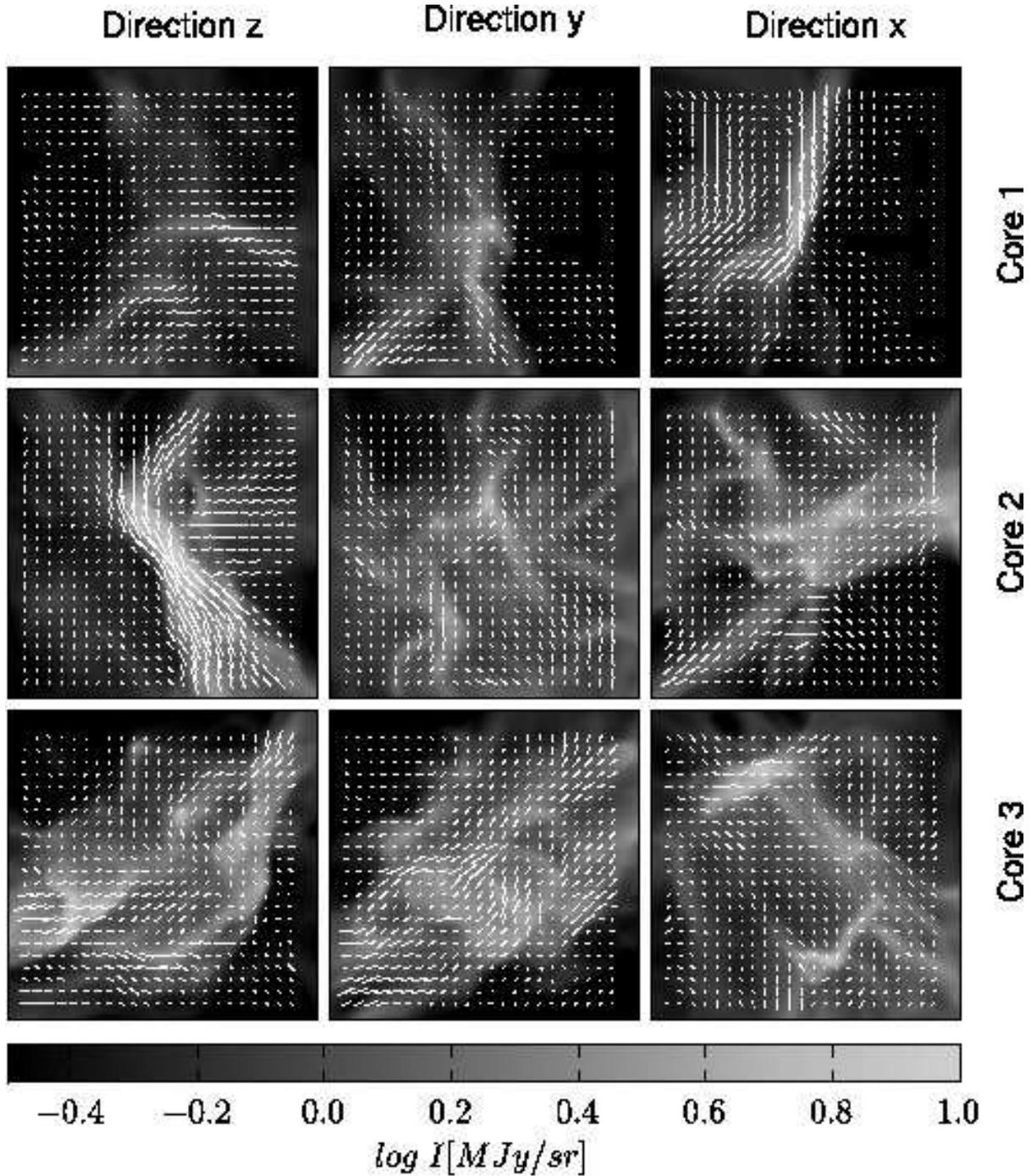} 
\caption{
Simulated polarization maps of the high resolution model at 353 GHz. Polarization vectors are drawn for every fifth pixel, the length corresponding to the polarized intensity. The background image shows the logarithm of the total intensity. The scaling of the background and the length of the vectors are the same in all frames, the longest vector being $0.11 \; {\rm MJy/sr}$.
} 
\label{3x3core}
\end{figure*}

\begin{figure*}
\centering 
\includegraphics[width=16cm]{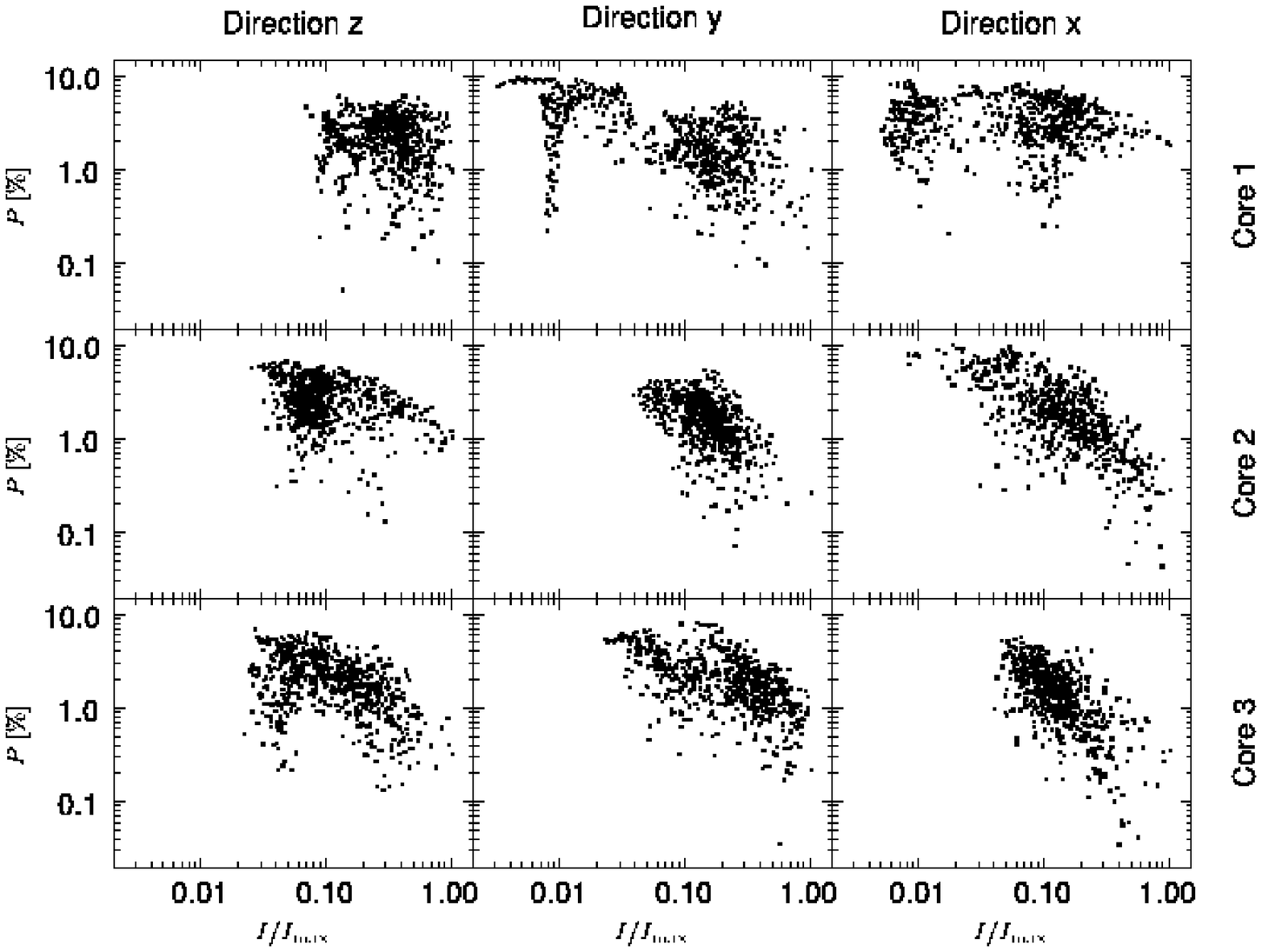} 
\caption{
The relation between polarization degree and total intensity for the polarization maps in Fig.~\ref{3x3core}. In order to avoid cluttering the plot, only every fifth map pixel along both horizontal and vertical direction is plotted.
} 
\label{poldeg}
\end{figure*}

\section{The model clouds}
\label{models}

We examine briefly one spherically symmetric cloud model for which the
density distribution is obtained from the Bonnor-Ebert solution of hydrostatic
equilibrium (Bonnor \cite{Bonnor1956}, Ebert \cite{Ebert1955}). The mass of the cloud is 3.7\,M$_{\sun}$ and, for a stability parameter value of $\xi=7.0$, the visual extinction to the cloud center is $A_{\rm V}\sim$15$^{\rm m}$. When the 
total opacity of the cloud is fixed the exact value of the parameter $\xi$ is
not important. For a given distance from the cloud surface, measured in
$A_{\rm V}$, the local intensity and anisotropy of the radiation field 
depend only weakly on the true shape of the density profile.

Our main interest lies in the study of inhomogeneous,
three-dimensional clouds. We revisit the model C discussed by Pelkonen et al. (\cite{Pelkonen2007}). It is based on the results of numerical simulations of highly supersonic magnetohydrodynamic turbulence, completed for a $128^3$ computational mesh with periodic boundary conditions (for details, see Pelkonen et al. \cite{Pelkonen2007}).

In addition, we study three cloud cores that are taken from a new simulation of
supersonic and super-Alfv\'enic MHD simulation. The MHD simulation was run on
a mesh of 1000$^3$ computational cells with the Stagger Code (Padoan et al.
\cite{Padoan2007}), re-meshed to 1024$^3$. In the used snapshot the rms sonic Mach number is 8.9
and the rms Alfv\'enic Mach number is 2.8, so that the turbulence is still
super-Alfv\'enic even with respect to the rms Alfv\'en velocity. From the
1024$^3$ volume, we selected three dense regions (`clumps'), each 128$^3$
cells in size (see Fig.~\ref{fig:1024_cloud_projection}). The average density
of the 1024$^3$ cloud was scaled to $n(\rm H_2)$=150\,cm$^{-3}$ and its linear
size was set to be 6\,pc. This corresponds to an average visual extinction of
$\sim$3$^{\rm m}$. In the selected core regions the visual extinction is higher,
reaching peak values of 10.9, 23.7, and 21.8 magnitudes in the three cores,
respectively. The radiative transfer is solved locally for the subgrids of $128^3$ grid points rather than the full $1024^3$ mesh.

\section{Results}
\label{results}

\subsection{Spherically symmetric cloud}

We start by looking at grain alignment in the case of the spherically
symmetric cloud described in Sect.~\ref{models}. Figure~\ref{1d_aniso_Av} shows the anisotropy factors at different wavelengths. The intensity of the short wavelength radiation drops sharply with $A_{\rm V}$, but the anisotropy increases since the remaining intensity comes preferentially from smaller and smaller space angles. At longer wavelengths, the dust emission starts to dominate over ISRF and the direction of the anisotropy is towards the surface, although the effect on effective $\gamma$ is small. At the center of the spherically symmetric cloud the anisotropy is by definition zero.

Figure~\ref{1d_min_align} shows the comparison of Eq.~\ref{eqaalg} and our results with constant $\gamma = 0.7$ and $\gamma$ calculated from the radiative transfer calculations. The number of grains larger than ~0.58 ${\rm \mu m}$ in Li \& Draine (\cite{Li2001}) dust model is negligible. Thus, Fig.~\ref{1d_min_align} shows that according to this calculation, there is essentially no alignment above 3 magnitudes, while in the Cho \& Lazarian model the limit is 5 magnitudes. Larger grains, if they exist, could be aligned by radiative torques at higher $A_{\rm V}$. In an inhomogeneous cloud there might also be some sightlines of lower $A_{\rm V}$, which produce more efficient radiative torques.

\subsection{Three-dimensional MHD models}

The dust polarization was studied in three-dimensional models described in Sect.~\ref{models}. Figure~\ref{old_new_emitted_353} shows the calculated polarization maps at 353 GHz for our old 3D model C, used in Pelkonen et al. (\cite{Pelkonen2007}). The polarization degree has been reduced everywhere, as expected from Fig.~\ref{1d_min_align}, but particularly in high density regions. This reduction in the polarization degree is more readily apparent in Fig.~\ref{old_new_corepoldeg}, where the $P/I$ slopes of most of the selected cores are steeper with the new calculations. This is because in a clumpy medium, starlight can penetrate the cloud from many different directions rather than in a clearly defined 'brightest' direction. Thus, the anisotropy will be averaged out somewhat, leading to weaker radiative torques and lower polarized emission. In the cores themselves the anisotropy is even less, so the difference with an invariant $\gamma = 0.7$ is greater, resulting in a steeper slope.

\begin{figure}
\centering 
\includegraphics[width=8cm]{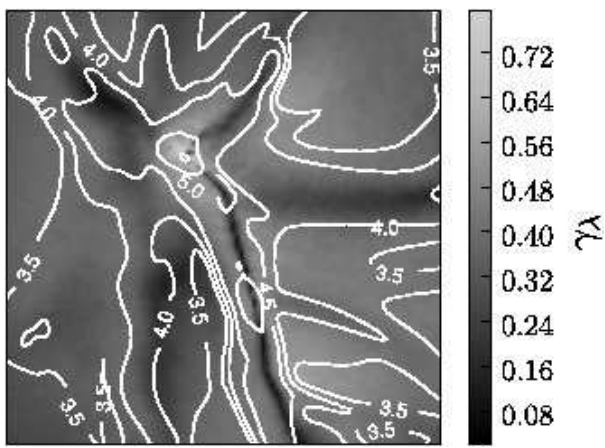} 
\caption{
The anisotropy factor $\gamma_{\lambda}$ at $\lambda = 0.55 \; {\rm nm}$ in an $xy$ slice. The slice is the mid-plane of the high density model for Core 2. The contours mark the density at $\log n_{H_2} = 3.5, 4.0, 4.5, 5.0, 5.5, 6.0$.
} 
\label{hiden_aniso}
\end{figure}

\begin{figure}
\centering 
\includegraphics[width=8cm]{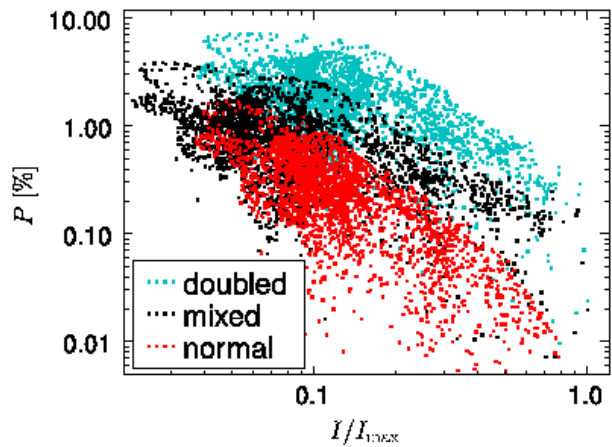} 
\caption{
The relation between polarization degree and total intensity for Core 2 (see Fig.~\ref{poldeg}) where the density has been multiplied by ten. The figure shows the polarization degree observed towards the $z$ direction. The polarized dust emission is calculated for doubled (approximate calculation, see text) and normal grain sizes, and for a mixed distribution where larger grains are found only in dense regions (see text).
} 
\label{ax2}
\end{figure}

\begin{figure*}
\centering 
\includegraphics[width=16cm]{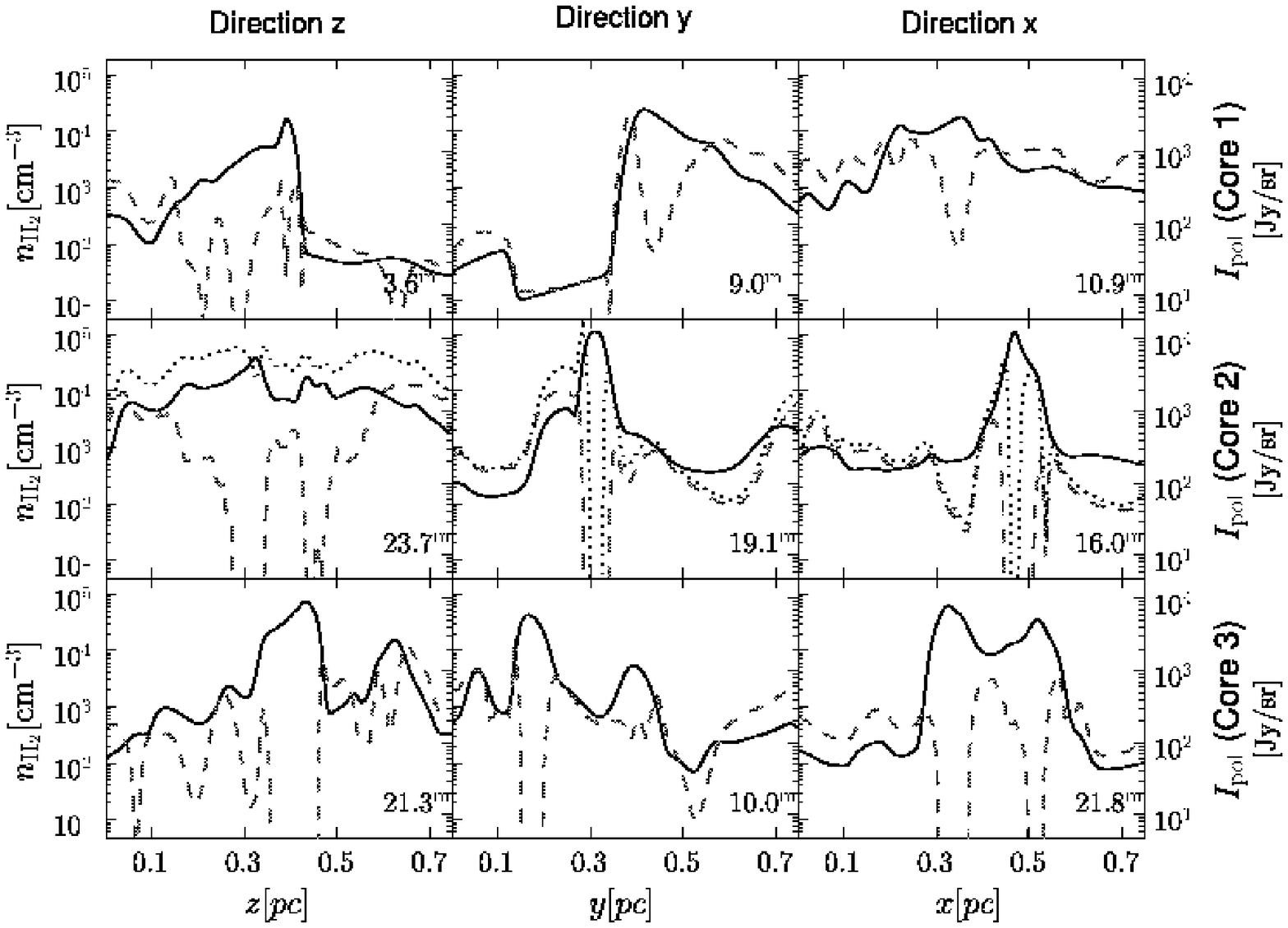} 
\caption{
The gas density, $n_{H_2}$, ({\it solid line}) and the polarized intensity ({\it dashed line}, at 353 GHz) along one line of sight through each core. The plots represents a line of sight through the intensity maximum of each map in Fig.~\ref{3x3core}. For Core 2, the polarized intensity is also calculated for doubled grain sizes ({\it dotted line}). The number in the lower right corner of each plot is the visual extinction along this line of sight.
} 
\label{LOS}
\end{figure*}

Figures~\ref{3x3core} and \ref{poldeg} show similar plots for our new high resolution model at 353 GHz. In Fig.~\ref{3x3core}, the vectors are drawn for polarized intensity, not for polarization degree, which is why long vectors are superimposed on the bright, high density regions. In Fig.~\ref{poldeg}, Core 1, when viewed from the $z$ direction, does not seem to show a clear decrease in the polarization degree, unlike the other examples. This is because Core 1 is less dense, and the maximum visual extinction through the core in the $z$ direction is only 3.9 magnitudes. This also explains why the other two relations for Core 1 are not as steep as for Cores 2 and 3, because radiation can penetrate the core much more easily.

\begin{figure}
\centering 
\includegraphics[width=8cm]{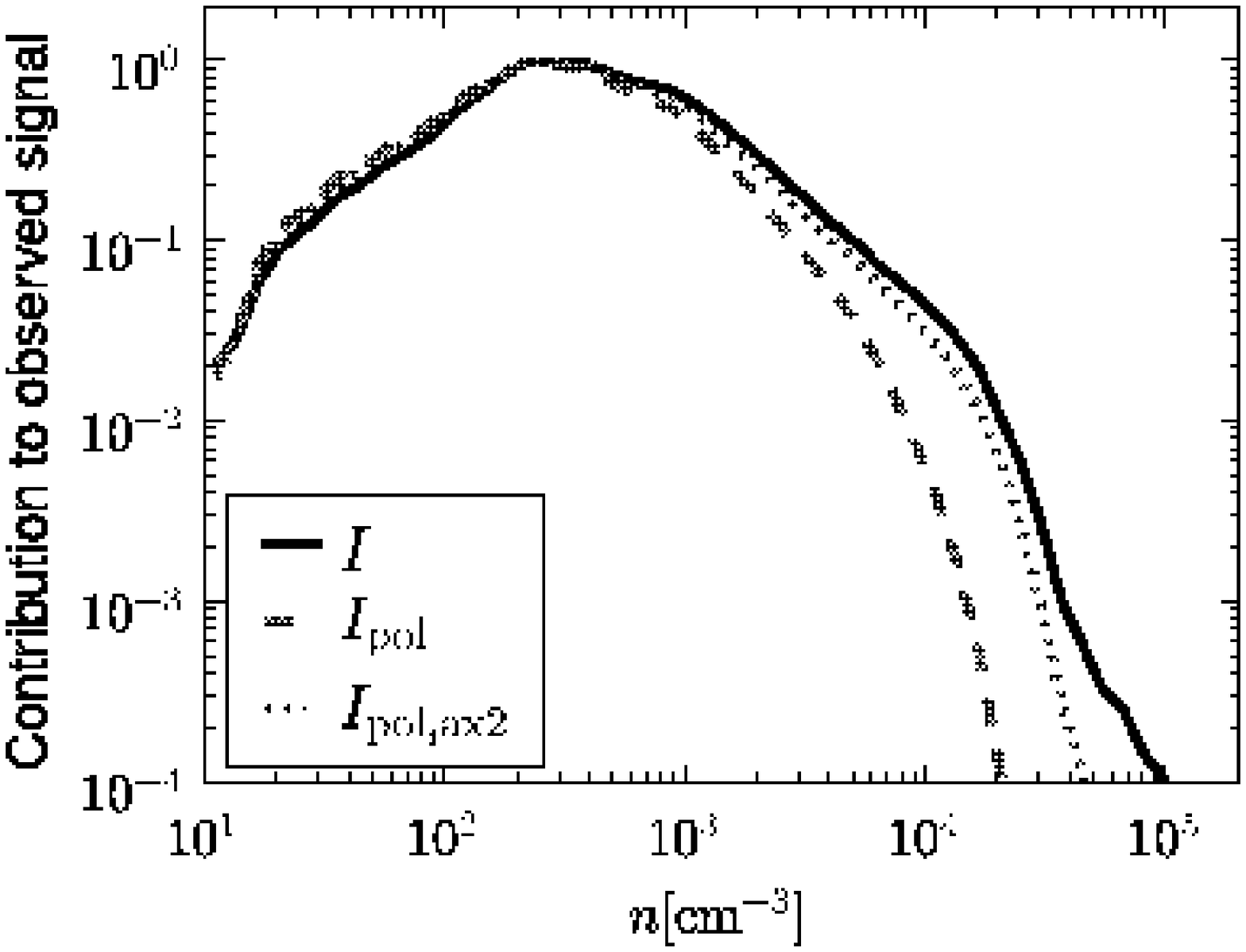} 
\caption{
A histogram showing the contribution of different density cells to the total intensity ({\it solid line}), and to the polarized intensity for normal grain sizes ({\it dashed line}) and doubled grain sizes ({\it dotted line}). This is for Core 2 and the local polarized emission as seen from the $z$ direction. The total and polarized intensity are shown as sums of all emission from the cells in that bin; thus the few high density cells have a low contribution to total intensity. All intensities are at 353 GHz.
} 
\label{C2_PI_histo}
\end{figure}

In dense clumps, grain coagulation can cause the grains to grow, shifting the size distribution to larger grain sizes. We took Core 2 and multiplied the density by ten to obtain a dense cloud. Figure~\ref{hiden_aniso} shows the anisotropy in a slice in the mid-plane, as well as density contours. In Fig.~\ref{ax2}, we show the polarization degree for three different dust size distributions for this high density model. We first calculated the grain alignment and dust emission using normal Li \& Draine (\cite{Li2001}) dust. The polarization degree is rather low and declines over the complete intensity interval. Secondly, we calculated the alignment by assuming that all our dust grains were double their original size. The shape of the radiative field inside the cloud should be closely the same, and we were only interested in the polarization degree and scaled intensity rather than absolute values of intensity. Thus, we used the same $a_{\rm alg}$ as for normal dust and simply used the doubled grain size distribution shown by Eq.~\ref{eqR}, resulting in more efficient polarization. There is a clear plateau in low intensity, before the grain size needed for alignment becomes larger and the polarization degree begins to drop. Finally, we investigated a model in which large grains appear only in dense regions. The two dust populations, one with normal grain size distribution and another with twice as large grain sizes, were calculated from the formulas:
\begin{eqnarray}
X_{d} & = & \frac{1}{1+e^{-\log n_5 / 0.3}},\\
X_{n} & = & 1 - X_{d},
\label{eqabu}
\end{eqnarray}
where $X_{d}$ and $X_{n}$ are the abundances of double sized and normal dust, and $n_5 = n_{\rm H} / 10^5 {\rm cm^{-3}}$.
The result, as expected, is between the two previous cases. There is a plateau, but it ends sooner than in the large grains case. However, towards the high intensity, the slope is becoming less steep, due to the grain growth in the densest clumps.

\begin{figure*}
\centering 
\includegraphics[width=16cm]{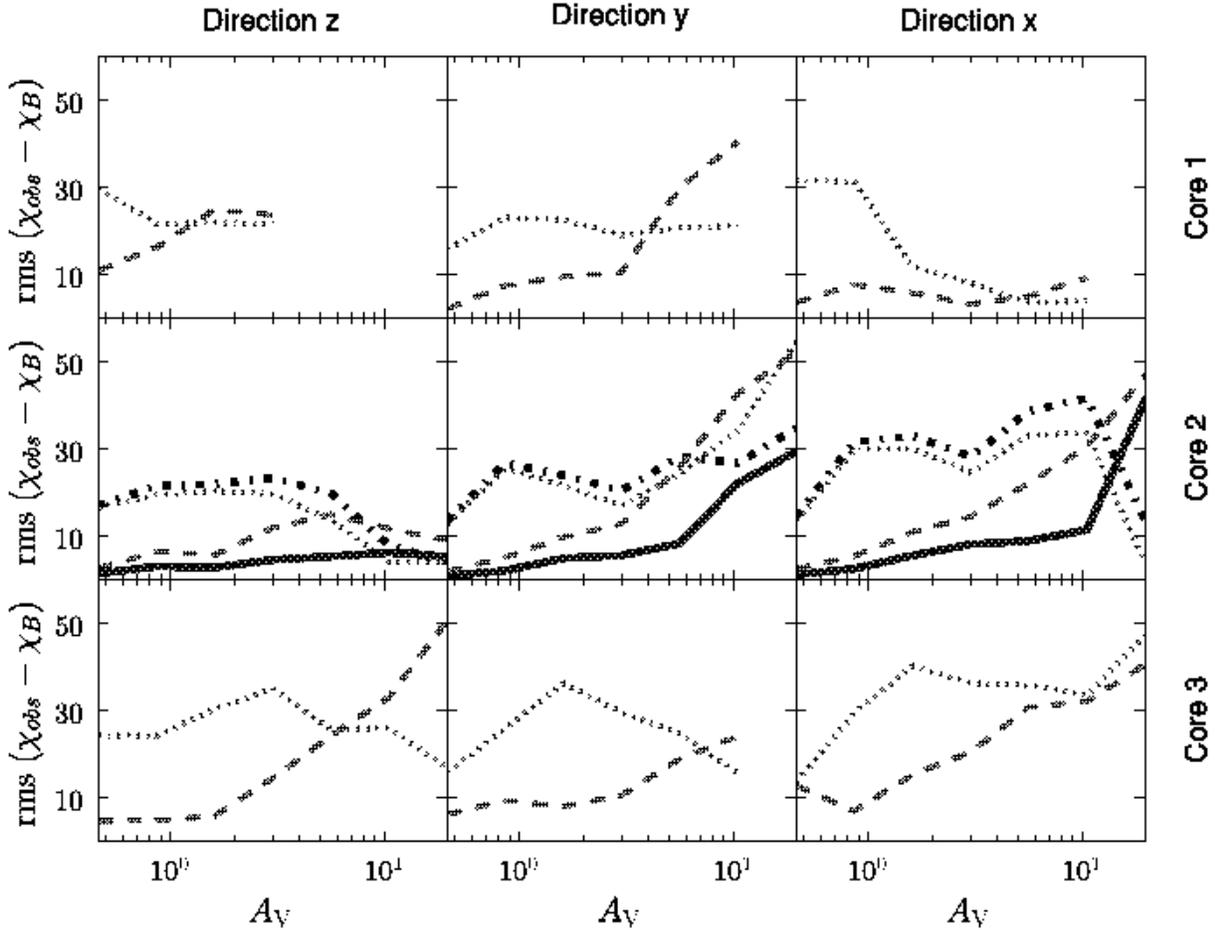} 
\caption{
The RMS difference between the observed polarization angles and the mass-averaged ({\it dashed line}) and volume-averaged ({\it dotted line}) magnetic field directions, for all cores and viewing directions. The observed polarization angles are calculated with normal Li \& Draine (2001) grain size distribution. For Core 2 polarization was also calculated with doubled grain sizes, and compared with mass-averaged (({\it thick solid line}) and volume-averaged ({\it thick dash-dot line}) magnetic field direction.
} 
\label{B_angle}
\end{figure*}

Where does the observed polarized emission originate? The thermal dust emission tends to trace the density structure. High density means more emitting dust and thus higher emission, although the temperature is anticorrelated with density and thus complicates the picture. However, Figs.~\ref{Raalg} and ~\ref{1d_min_align} hint that above a couple of $A_{\rm V}$ only a fraction of the grains are aligned, and thus the polarization degree drops sharply. Figure~\ref{LOS} shows the polarized intensity and the gas density in each cell along lines of sight through each of the cores as seen from three different viewing directions. The effect of shadowing by even modest density structures is striking, as is the absence or reduction in the polarized emission from the densest structures. In Core 2, the polarized emission is also plotted for the doubled grain size. In this case, the polarized emission originates deeper inside the core. However, because of the high local density, the density maximums in the $y$ and $x$ directions still do not have aligned grains. The $z$ direction exhibits significant polarized emission from the density maximum even though the visual extinction along that line of sight is greater. In Core 1, the magnetic field is oriented almost towards the observer in the $z$ direction, which is why the polarization is weakened.

Figure~\ref{C2_PI_histo} shows the histogram of both the intensity and the fraction of polarized intensity in different density bins for Core 2 seen along the $z$ direction. Due to the summing of the contribution of the individual cells in each bin, we need to compare the total intensity contribution with the polarized emission contribution. The high density cells contribute less and less to the polarized emission, while the contribution of the small density cells is enhanced. With doubled grain sizes, the more efficient grain alignment is clearly evident because the polarized emission still comes from higher density regions, before finally falling, too.

Do polarization vectors trace the magnetic field with any degree of certainty? Above a threshold, which depends on the local density as well as the geometry of the cloud, the radiative torque alignment is no longer capable of organizing the grain orientation. Thus the emission from shielded region is unpolarized, giving the observer no information about the magnetic field within it. 

Figure~\ref{B_angle} shows the rms errors between the observed polarization angle, $\chi_{\rm obs}$, and the volume- and the mass-averaged magnetic field angles, $\chi_{B,\rm vol}$ and $\chi_{B,\rm mass}$. For low $A_{\rm V}$, $\chi_{B,\rm mass}$ and $\chi_{\rm obs}$ agree well, because there are no dense clumps along the line of sight where the radiative torque alignment would be inefficient. Temperature differences between warm diffuse regions and cold denser regions do introduce some discrepancies caused by the different weighting in emission and in density. For higher $A_{\rm V}$, the situation changes; $\chi_{B,\rm mass}$ traces the magnetic field directions in the dense clumps, while the influence of dense clumps on $\chi_{\rm obs}$ is reduced by the weakened radiative torque alignment. This is because the radiative torque alignment continues to be effective on the larger grains, and the dense clumps continue to contribute to the polarized emission. Any emission or density variations are not taken into account by $\chi_{B,\rm vol}$, and thus the difference between observed and volume-averaged angles depends greatly on the density profile and the local magnetic field orientation. If the density profile is flat or the magnetic field is ordered apart from the densest clumps, then $\chi_{B,\rm vol} \sim \chi_{\rm obs}$.

At the highest $A_{\rm V}$, there are only a few independent structures (see Fig~\ref{LOS}), and the different magnetic field geometries play a crucial role. Taking Core 2 as an example, there is  one structure along the $z$-axis. In that structure, two clumps along the line of sight have their magnetic fields aligned anti-parallel. Their combined signal is mostly unpolarized, and thus $\chi_{\rm obs} \sim \chi_{B,\rm mass} \sim \chi_{B,\rm vol}$. That is why the rms remains flat even to high $A_{\rm V}$. The same would happen if both clumps had a magnetic field direction close to $\chi_{B,\rm vol}$, in which case the contribution of the clumps - whether total ($\chi_{B,\rm mass}$) or partial ($\chi_{\rm obs}$) - would only reinforce the orientation from the diffuse regions. Along the $y$-axis, the density profile is quite varied, as is the magnetic field orientation, and thus none of the angles agree at high $A_{\rm V}$. Finally, along the $x$-axis, direction of the mass-averaged magnetic field is determined mostly by one high density clump, where the magnetic field direction spins around. Thus, $\chi_{\rm obs} \neq \chi_{B,\rm mass}$ but  $\chi_{\rm obs} \sim \chi_{B,\rm vol}$. As always with polarization, one cannot ignore the geometrical effects.

From Fig.~\ref{B_angle}, it can be seen that an rms difference of 10-30 degrees is common for sightlines of higher extinction than a few $A_{\rm V}$. This difference arises because the grains are not aligned in the dense regions of the cloud, and thus the dense regions do not contribute to the observed polarization. This may cause problems for the Chandrasekhar-Fermi method (Chandrasekhar \& Fermi \cite{Chandra1953}) and in attempts to infer core formation processes based on the relative alignment of B and core geometry.

\begin{figure}
\centering 
\includegraphics[width=8cm]{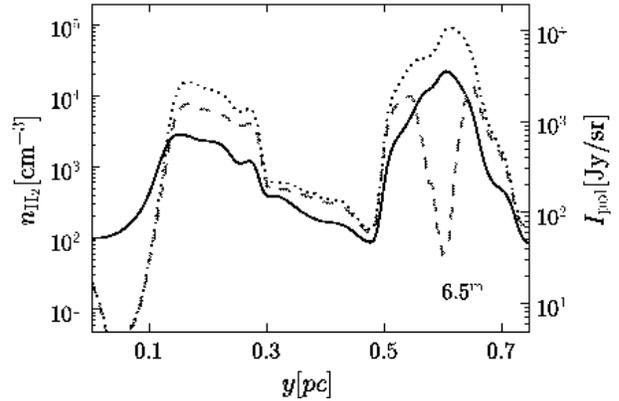} 
\caption{
The gas density, $n_{H_2}$, ({\it solid line}) and the polarized intensity ({\it dashed line}) along one line of sight through Core 2. The polarized intensity is also calculated for doubled grain size ({\it dotted line}). The number in the lower right corner is the visual extinction.
} 
\label{LOS_example}
\end{figure}

Despite the above caveat, the $\chi_{\rm obs}$ and $\chi_{B,\rm mass}$ do seem to agree until the extinction exceeds a couple of $A_{\rm V}$. If grain size increases (Core 2, solid line), $\chi_{\rm obs}$ and $\chi_{B,\rm mass}$ can remain in a good agreement even to $A_{\rm V} \sim 10^{\rm m}$, as predicted by Cho \& Lazarian (\cite{cholazarian}). This is because the radiative torque alignment continues to be effective on the larger grains, and the dense clumps continue to dominate the polarized emission. Figure~\ref{LOS_example} shows an example along one line of sight where the doubled grain size allows the dense clump to be probed.

\begin{figure}
\centering 
\includegraphics[width=8cm]{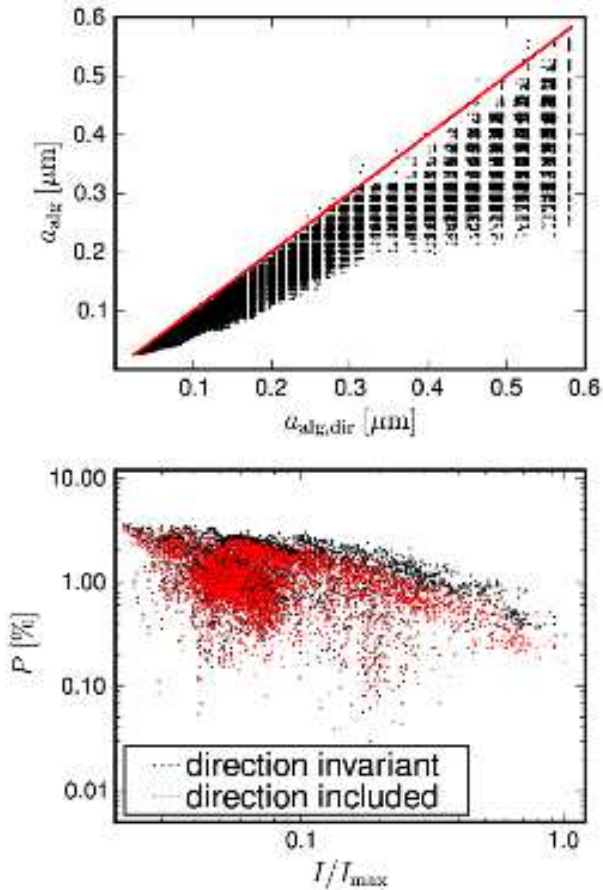} 
\caption{
{\it Upper panel:} The aligned grain sizes for direction-dependent radiative torques, $a_{\rm alg,dir}$, and direction-invariant radiative torques, $a_{\rm alg}$. This is for Core 2, taking into account only the anisotropic component. The solid line shows where $a_{\rm alg,dir} = a_{\rm alg}$. The gaps in the grain sizes is due to the size grid used in the calculations. {\it Lower panel:} The relation between polarization degree and total intensity calculated using the above grain sizes and seen along the $z$-axis.
} 
\label{dir_comp}
\end{figure}

Finally, we examine the effect of the dependence of the radiative torque efficiency on the angle between the radiation field and the magnetic field. We limit the study to the anisotropic component of the radiation field. In Fig.~\ref{dir_comp}, we can see the clear need for larger grains because of the weakening of the radiative torques at larger angles. Some isolated cells show the opposite effect, where the summed radiative torque is actually stronger due to less efficient opposing torques. Usually, there is a drop of a factor of two in the polarization degree of the denser regions.

The weakening of the radiative torques means that we miss more of the polarized emission from dense clumps in the clouds. This limits our ability to trace the mass-averaged magnetic field direction along sightlines of high $A_{\rm V}$. In the normal grain size case, our ability to trace the magnetic field dropped from $A_{\rm V} \sim 3^{\rm m}$ to $1^{\rm m}$. However, this is a complex issue. In Fig.~\ref{B_angle}, we can see examples where magnetic field is traceable over $A_{\rm V} = 10^{\rm m}$ even for normal grains, because of the orientation of the magnetic field. Also, the $A_{\rm V}$ limit depends on the direction the core is viewed from. For example, Core 2 gives quite different results for each different viewing direction. Changing our alignment parameter, $\omega_{\rm rad}/\omega_{T} =  5$, to 3 (see Discussion) would also improve the alignment, helping to counteract some of the weakening of the radiative torques. Grain growth is of critical importance. Based on our current results, we conclude that $A_{\rm V} = 10^{\rm m}$ is likely to be an upper limit, and polarized emission provides reliable data on magnetic fields only up to a few magnitudes in $A_{\rm V}$.

\section{Discussion}

One critical parameter for grain alignment by radiative torques is the grain size. Figure~\ref{B_angle} shows that in the case of a diffuse-medium dust model the observed polarization vector no longer traces the mass-weighted magnetic field direction in clumps of extinction higher than a few $A_{\rm V}$. However, if the grain size doubles, the polarization vectors can carry information about far denser clumps. This is highly dependent on the geometry of the cloud, as can be seen in Core 2, where, along the $y$-axis, we reach only $A_{\rm V} \sim 7^{\rm m}$, while along the $x$-axis we reach $\sim 10^{\rm m}$. A doubling of the grain size is not improbable: the sub-mm emissivity of dust has been observed to increase significantly in clouds that have a visual extinction of only a few magnitudes (Bernard et al. \cite{Bernard1999}; Stepnik et al. \cite{Stepnik2003}). This has been interpreted as evidence of grain growth and also of the appearance of dust aggregates and fluffy grains of low packing density. Whittet (\cite{Whittet2007}) found evidence in a globule of dust growth to mean sizes 50 \% - 100 \% larger than in the diffuse ISM, even along a sightline of $A_{\rm V} = 1.1$.

For radiative torques the important question is the growth of the largest grains. Ossenkopf \& Henning (\cite{Ossenkopf1994}) presented several models of grain growth with thin and thick ice mantles, and coagulation. They were able to explain the observed changes in dust opacity. In their model of thick ice mantles, applicable to cold dense cores, the ice volume fraction was a factor of 4.5 above the original dust volume, causing a significant change in the size distribution. But since the ice mantle thickness is almost constant irrespective of the grain size, most of this increase in size happens in the smaller grains. Grain coagulation is needed to make the large grains larger, and this may take longer than the free fall time of a cloud (Ossenkopf \& Henning \cite{Ossenkopf1994}; Chakrabarti \& McKee \cite{chmc05}). Hence, the magnetic field within the youngest cores may not be accessible, while it could be probed inside older cores and filaments that survived significantly longer than their free-fall time.

Many observations of the dust polarization at sub-mm wavelengths focus on clouds that are rather luminous in dust emission (e.g. Davis et al. \cite{Davis2000}; Matthews \& Wilson \cite{Matthews2000}, \cite{Matthews2002}; Henning et al. \cite{Henning2001}; Wolf et al. \cite{Wolf2003}; Lai et al. \cite{Lai2003}; Crutcher \cite{Crutcher2004}), many with internal sources. Thus, the comparison of our results derived from a MHD modeling with cold dust and without internal sources is not straightforward. However, it is interesting to note that if the intensities are normalized by the maximum intensity of a given core, the modeled $P/I$ -relations show qualitatively similar behavior to the observed ones, dropping from ~10\% to ~1\% in polarization degree. 

Cho \& Lazarian (\cite{cholazarian}) chose their alignment parameter, $\omega_{\rm rad}/\omega_{T} =  5$, and derived Eq.~\ref{eqaalg} from that assumption. Lazarian \& Hoang (\cite{Lazarian2007}) presented a simple toy model of a helical grain to study the basic properties of radiative torque alignment, and in their later work (Hoang \& Lazarian \cite{Hoang2008}) found that when $\omega_{\rm rad}/\omega_{T} =  3$, a significant fraction of the grains are aligned. Bethell et al. (\cite{bethell}) decided to use $\omega_{\rm rad}/\omega_{T} =  3$ in their own study of the anisotropy of the radiation field inside a clumpy cloud. We chose to use the higher value to be able to compare the results to our previous paper (Pelkonen et al. \cite{Pelkonen2007}). The lower alignment parameter leads to a more effective alignment of grains. This does not dramatically change the results of this study, since, as seen in Fig~\ref{fig2}, Eq.~\ref{eq5} is very sensitive to the grain size. As a result, we need only slightly larger grains than Bethell et al. (\cite{bethell}).

Direct comparison with the results of Bethell et al. (\cite{bethell}) is further complicated by the different cloud models and grain size distributions. However, with their mass-averaged anisotropy factor $\gamma = 0.34$, we would not expect isotropic radiative torques to play any significant role. In our high-density model that showed similarly high anisotropy, isotropic contribution was negligible. In the low-density case, anisotropy was low and thus isotropic contribution was noticeable as a small enhancement of polarization degree. Despite the setup differences with the study of Bethell et al. (\cite{bethell}), we have obtained qualitatively similar results and hence agreement with their conclusions. However, we would like to add one note of caution to their statement about invariant anisotropy having only a moderate effect on the polarization degree. When looking at the emergent polarized emission along a line of sight, it is obvious that often the very densest clumps do not contribute at all. This is due partly to the weakening of the radiation field, but also to the smaller anisotropy inside the core. Adopting an invariant anisotropy would decrease the contribution of the shell around the core, where the anisotropy is highest, and might cause the grains in the dense core to be aligned. Hence, while the big picture of the $P/I$ relation might look similar, the magnetic field that is sampled in the model might be quite different.

\section{Conclusions}

We have examined the anisotropy of radiation in spherically symmetric cloud models and in inhomogeneous 3D cloud models. In our 1D cloud model, the anisotropy was well below 0.7 (Fig.~\ref{1d_aniso_Av}), resulting in a weaker radiative torque alignment (Fig.~\ref{1d_min_align}). 

In a 3D cloud, the line of sight $A_{\rm V}$ can be high, while the effective $A_{\rm V}$ can be low due to inhomogeneity. These low $A_{\rm V}$ sightlines can reduce the anisotropy because of averaging and thus reduce the polarized signal, as also noted previously by Bethell et al. (\cite{bethell}). Total anisotropy is sensitive to the geometry of the cloud, but also to the $A_{\rm V}$ of the cloud. The $P/I$ relations show a qualitatively similar behavior to the observed ones.

The 3D model and already the 1D case show that radiative torque alignment does not operate well inside cloud cores, where $A_{\rm V}$ is larger than a few magnitudes. Thus, we should be careful not to interpret the polarization that we see from a dense core as originating in the core, but rather from the more diffuse regions on the line of sight, where the ISRF is able to align the grains via radiative torques (Fig.~\ref{LOS}). These results suggest that one must be careful with models of core formation. We intend to study the impact on the Chandrasekhar-Fermi method in future papers.

The growth of grains in the clumps could result in a more efficient alignment of grains even in the denser clumps, allowing them to continue contributing to the observed polarized emission. Thus, if grain growth occurs in the clumps, the observed polarization vectors may trace the magnetic field lines possibly up to $A_{\rm V} \sim 10$ magnitudes, in agreement with the prediction in Cho \& Lazarian (\cite{cholazarian}) and the results of Bethell et al. (\cite{bethell}). 
However, prestellar cores are often found to evolve on dynamical timescales, 
suggesting that larger grains would not have time to coagulate, and polarization vectors would not trace the magnetic field within the cores. Furthermore, using radiative torques that account for the angle between the radiation direction and the magnetic field, the polarization is again reduced. As a result, dense cores would need even larger grains for their inner magnetic fields to be detectable. Even with grain sizes at twice the size, we are likely to trace magnetic fields only up to a few magnitudes in $A_{\rm V}$.

\begin{acknowledgements}

V.-M.P. and M.J. acknowledge the support of the Academy of Finland Grants no.
206049, 115056, 107701 and 124620. P.P. was partially supported by the NASA ATP grant NNG056601G and the NSF grant AST-0507768.

\end{acknowledgements}

\end{document}